\documentclass[aps,pre,preprint,groupedaddress]{revtex4-1}

\usepackage{graphicx}
\usepackage[dvips]{color}
\usepackage{natbib}
\usepackage{color}

\newcommand\Tr{\mbox{\textit{Tr}}}  % Transmission coefficient

\begin{document}

%\preprint{}

\title{Particle-scale structure in frozen colloidal suspensions\\ from small angle x-ray scattering}


\author{Melissa Spannuth}
\email[]{melissa.spannuth@gmail.com}
\thanks{Present address: Department of Chemical and Biomolecular Engineering, University of Houston, Houston, TX 77204, USA}

%\thanks{}
\affiliation{Department of Geology and Geophysics, Yale University, New Haven, CT 06520, USA}

\author{S. G. J. Mochrie}
\affiliation{Department of Physics, Yale University, New Haven, CT 06520, USA}

\author{S. S. L. Peppin}
\affiliation{OCCAM, Mathematical Institute, University of Oxford, St. Giles Road, Oxford OX1 3LB, UK}

\author{J. S. Wettlaufer}
\affiliation{Department of Geology and Geophysics, Yale University, New Haven, CT 06520, USA}
\affiliation{Department of Physics, Yale University, New Haven, CT 06520, USA}
\affiliation{Program in Applied Mathematics, Yale University, New Haven, CT 06520, USA}


\date{\today}

\begin{abstract}
During directional solidification of the solvent in a colloidal suspension, the colloidal particles segregate from the growing solid, forming high-particle-density regions with structure on a hierarchy of length scales ranging from that of the particle-scale packing to the large-scale spacing between these regions. Previous work has concentrated mostly on the medium- to large-length scale structure, as it is the most accessible and thought to be more technologically relevant. However, the packing of the colloids at the particle-scale is an important component not only in theoretical descriptions of the segregation process, but also to the utility of freeze-cast materials for new applications. Here we present the results of experiments in which we investigated this structure across a wide range of length scales using a combination of small angle x-ray scattering and direct optical imaging. As expected, during freezing the particles were concentrated into regions between ice dendrites forming a microscopic pattern of high- and low-particle-density regions. X-ray scattering indicates that the particles in the high density regions were so closely packed as to be touching. However, the arrangement of the particles does not conform to that predicted by standard inter-particle pair potentials, suggesting that the particle packing induced by freezing differs from that formed during equilibrium densification processes.

\end{abstract}

\pacs{82.70.Dd,64.75.Xc}
%\keywords{}

\maketitle

% **** INTRODUCTION **** %

Interest in directional solidification, or freeze-casting, of suspensions of particles has surged recently owing to the relative versatility, simplicity and cost-efficiency of this process for fabricating complex composite materials \cite{deville2008}. This method has been used to create materials for applications such as tissue scaffolds \cite{fu2008}, biomimetic materials \cite{deville2006}, photonic structures \cite{kim2009}, and metal-matrix composites \cite{wilde2000}. In addition, directional solidification has been shown to be effective for water purification \cite{gay2002} and occurs naturally when the ground freezes \cite{dash2006}. In all of these cases, the segregation of particles from the growing solid and the consequent increase of particle concentration in the fluid regions are paramount. In particular, the structure of the regions of segregated particles is important for performance of the material in many applications. This structure occurs on a variety of length scales from the relatively large scale of individual regions of segregated particles to the single particle scale of the packing density of segregated particles. Although most research has focused on the large-scale structure, the particle-scale structure is key to understanding the particle rejection behavior and hence predicting the large-scale structure.

In a very dilute suspension, rejection of single particles from a solidfication front is well understood as resulting from fluid flow into the premelted film that separates the particles from the growing solid (e.g., references \cite{uhlmann1964,azouni1990,rempel1999}). In non-dilute suspensions, the same fundamental rejection mechanism is responsible for particle segregation during solidification, but the comprehensive interaction between the growing solid and the large number of particles found in non-dilute suspensions is not well understood. Conceptually, rejection increases the particle concentration in the fluid until the concentration reaches a threshold. Further particle rejection is untenable and the solidification front either becomes unstable or engulfs particles, or both \cite{peppin2007}. The morphology of the instability and the mode of particle incorporation creates macro- and microscopic patterns of high- and low-particle-density regions. Depending upon the freezing conditions, commonly observed patterns include, among others, lamellae oriented parallel or perpendicular to the solidification direction, branching or hexagonal networks of nearly pure solid, and seemingly disordered crack-like patterns (e.g., references \cite{taber1929,watanabe2000,zhang2005,peppin2007,peppin2008,deville2009,peppin2009}). Similarities between these patterns and those formed during drying of colloidal suspensions (e.g., reference \cite{allain1995}) or jamming of suspensions flowing through constrictions (e.g., references \cite{haw2004,campbell2010}) suggest that the physics underlying the colloid behavior may be similar as well, though the driving forces in each case differ. Thus, knowledge gained from studying structures in freezing colloidal suspensions may be applicable to dense colloidal suspensions in diverse circumstances.

Presently, there is no theory that can fully predict the morphology or detailed characteristics of the patterns that form. However, a continuum approach analogous to that describing binary alloy solidification has been successful in predicting the transition from particle pushing to particle capture \cite{peppin2007,peppin2008,deville2009}. This description requires information about the particles near the freezing front, such as the packing density and rate of diffusion in the suspension. Although these quantities have been modelled assuming that the particles behave as hard spheres, it is unknown whether this equilibrium approach to the statistical mechanics of the particles is accurate, or whether the forces associated with the solid growth and concomitant fluid flow affect the particle behavior. Furthermore, a hard sphere pair potential is not a good approximation of the inter-particle interactions for many systems of interest. Therefore, it is important to understand the particle-scale structure and behavior in solidifying colloidal suspensions.

This type of information is difficult to obtain experimentally because the particle concentrations and materials typically involved make the suspensions opaque to visible light. In addition, the particles are often too small to observe individually and the structures that form are three-dimensional. As a result, most studies involve postmortem analysis of samples after sublimation of the solid and sintering or other fixing of the particle structure (e.g., references \cite{zhang2005,deville2006,shanti2006,fu2008,waschkies2009,kim2009}). This gives only a two-dimensional view of the three-dimensional structure, provides only static information about the final particle arrangement, and may be skewed by modification of the structure during sublimation and sintering \cite{deville2009b}.

A couple of experiments have overcome some of these difficulties by using either a very thin sample cell and transparent materials \cite{sekhar1991}, or applying x-ray techniques (radiography and tomography) to thicker samples \cite{deville2009b}. The thin sample chamber produces a quasi-two-dimensional system that can be observed with visible light microscopy for sufficiently low particle concentrations, while x-ray techniques can probe inside visibly opaque samples. X-ray tomography can even provide a full three-dimensional reconstruction of the samples. All allow samples to be viewed during the freezing process, though the long acquisition time for tomography allows only relatively slow solidification rates \cite{deville2009}. Improved x-ray tomography may relax this restriction \cite{weitkamp2010}. Yet, none of these techniques provide information about the particle-scale structure of the samples. In order to obtain this information, we used small angle x-ray scattering (SAXS), which provides a Fourier-space representation of the mass distribution within the samples on the scale of one to several times the particle radius.

Here we present the results of a combined x-ray scattering and direct imaging study. Our experiments benefit from the relative simplicity of a thin sample chamber, which allows sufficient light transmission to produce direct images of the samples. The images provide a basis for interpreting the SAXS intensity data collected before freezing, after melting, and while the samples were frozen. Most importantly, while frozen the data exhibit features related to the structure of the regions of segregated particles that formed during freezing. In particular, we find that the particles are very densely packed, even touching, and their arrangement does not conform to any predicted by standard models of inter-particle interactions. Therefore, the freezing process must cause particles to pack together in an unusual manner, possibly by creating inter-particle pressures that cannot be attained in the unfrozen solutions. This is an important point that must eventually be accounted for in solidification models, but more generally it raises questions about the arrangement of particles in dense suspensions under external forcing.

% **** MATERIALS AND METHODS **** %

\section{Materials and Methods}

For our x-ray scattering experiments, we used solutions of colloidal silica spheres dispersed in deionized water contained within a specially designed thin, transparent sample chamber. The choice of materials and the experimental setup were each tailored to the specific requirements of the x-ray scattering experiment. This section provides the details of the samples, sample cell, and other aspects of the procedures used in the experiments.

\subsection{Materials}
\label{materials}

Our samples consisted of colloidal silica spheres (Bangs Labs) with radii of about $32$ nm and polydispersity of about $18$\%, as determined from scanning electron micrographs and SAXS data (discussed below). The particles were stabilized against aggregation by surface-induced ionization. We modified the as-received solutions by centrifuging to sediment the particles and then replacing the supernatant with deionized water (Fisher Scientific deionized, ultrafiltered; resistivity $0.5$ M$\Omega/$cm) in order to remove as much as possible of the ionic species (NaOH) added as a stabilizer by the manufacturer, though the final solutions likely still contained some small amount of free ions \cite{mythesis}. Removal of the dissolved ions is important because they complicate interpretation of the experiments by affecting the stability of the solidification front \cite{mullins1964}, depressing the melting temperature of the solution \cite{dash2006}, and congregating in large melt pockets long before bulk melting occurs \cite{mythesis}. Although removing the dissolved ions could destabilize the colloids and lead to aggregation, we did not observe any indications of this prior to freezing the solutions.

During centrifuging, we also adjusted the particle volume fraction of the solutions to $\phi_{HS} \approx 0.07$--$0.08$, where $\phi_{HS}$ is the volume fraction of equivalent hard spheres. This was estimated from the manufacturer's stated volume fraction and the amount of solvent removed, and was verified by the SAXS data assuming hard sphere interactions (discussed below). The actual particle volume fraction based on the physical particle radius was $\phi \approx 0.02$.

\subsection{Sample Cell}

\begin{figure}[tp]
\begin{center}
\includegraphics[]{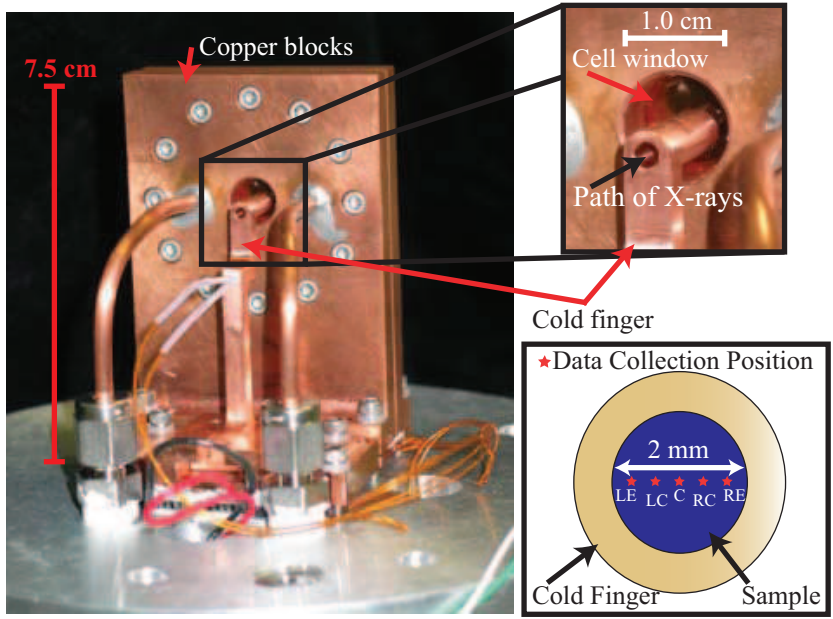} 
\caption[The sample chamber]{The image on the left shows the entire sample cell with the sample chamber and cold finger tip enlarged in the top right hand corner. In the lower right hand corner, the schematic diagram shows a plan-view of the cold finger tip with the approximate locations of x-ray data collection (LE = left edge, LC = left center, C = center, RC = right center, and RE = right edge).}
\label{samplechamber}
\end{center}
\end{figure}

The sample chamber within the cell was formed by sandwiching an approximately $400 \, \mu$m thick aluminum washer between two copper blocks. Circular pieces of thin polyimide film (Kapton) were epoxied across circular holes on each block to form the viewing area (Fig. \ref{samplechamber}). A thermoelectric cooling device (TEC, or Peltier cooler) in contact with the copper blocks controlled their temperature. A second TEC controlled the temperature of a copper arm (the ``cold finger'') that made thermal contact with the sample through physical contact with the outside of one of the windows. The cold finger had a cylindrical tip with inner diameter $2$ mm and outer diameter $4$ mm. By maintaining the temperature of the blocks above $0^{\circ}$C while that of the cold finger was lowered below $0^{\circ}$C, we created a nearly isothermal region within the cold finger inner diameter and a temperature gradient region between the cold finger outer diameter and the blocks. This allowed continuous contact with a reservoir of unfrozen solution, which helped alleviate pressure build-up during freezing and due to frost heaving when frozen \cite{wilen1995}. The temperature control system and calibrated platinum resistance thermometric devices (Pt RTD's) provided $\pm 0.001^{\circ}$C precision and $\pm 0.05^{\circ}$C accuracy in temperature measurement, as well as temperature stability of $\pm 0.001^{\circ}$C over $10$ minutes. Finally, the actual thickness of the sample chamber varied between about $200 \, \mu$m and $400 \, \mu$m due to the flexibility of the windows combined with manual positioning of the cold finger abutting one window. 

\subsection{Procedure}

The x-ray scattering experiments were performed at beam line 8-ID of the Advanced Photon Source at Argonne National Laboratory. Details of the beam line are provided by references \cite{lumma2000pre} and \cite{falus2004}, but we will summarize the important aspects in this section along with the details of our particular experiment at this beam line.

For the x-ray experiments, the sample cell described above was placed in the beam line, which was evacuated to about $10^{-2}$ torr. Evacuating the beam line minimizes stray scattering of the x-ray beam from air or water vapor as it approaches the sample and then as the scattered x-rays travel to the detector. The fluid inside the sample chamber remained at atmospheric pressure because it was connected to the ambient atmosphere via the fill lines.

In four separate trials, we investigated four different samples, each prepared in the same manner and labeled samples $1$--$4$ in the results below. Each of the samples was frozen by lowering the cold finger temperature to around $-30^{\circ}$C, while the temperature of the blocks was maintained at a constant $1^{\circ}$C throughout all experiments. The samples cooled at rates up to $1^{\circ}$C/s at higher temperatures and nearly $0.25^{\circ}$C$/$s at lower temperatures. Ice typically nucleated between $-20^{\circ}$C and $-30^{\circ}$C, manifested by a slight change in the rate of decrease of the temperature due to the release of latent heat. After freezing, we studied the samples at temperatures between $-2^{\circ}$C and $0^{\circ}$C with intervals as small as $0.05^{\circ}$C, always increasing the temperature over time. Thus, temperature increased as the sample age increased, though not continuously and not at precisely the same rate in all experiments. As a result, effects due to the increasing temperature and aging of the samples are convoluted in our experiments.

We acquired x-ray scattering data at many temperatures before freezing, immediately after freezing, and as the temperature was increased towards $0^{\circ}$C. We could not acquire data during freezing due to the unpredictable timing of ice nucleation and the speed of ice growth in the highly supercooled suspension. At each temperature, the x-ray beam was directed through the inner diameter of the cold finger and positioned at each of five different locations across this region as shown in Fig. \ref{samplechamber}. Thus, the x-ray experiments interrogated several parts of the isothermal region of the samples.

The x-rays we used had an energy of approximately $7.4$ keV for a wavelength of about $0.17$ nm. The beam cross-section was roughly $20 \, \mu$m by $20 \, \mu$m with a total incident flux of approximately $4 \times 10^{9}$ photons$/$s. For comparison, the cell thickness is several hundred $\mu$m and the particle radius is only $0.032 \, \mu$m, so there are millions of particles in the scattering volume.

The scattered x-rays were collected by a charge-coupled device (CCD) camera, described in reference \cite{falus2004}. The CCD detector was exposed to scattered x-rays for $0.015$ s per frame. To form a data set, a total of $500$ frames were collected over about $110$ s. During the readout time between frames and whenever data were not being acquired, the sample was blocked from x-ray illumination to limit radiation damage, which may induce melting \citep{schoder2009}.

Each frame in a particular data set was analyzed to create false color images of the scattered intensity. We verified that the scattering pattern was isotropic and did not change significantly while acquiring a set of images. Therefore, the images could be averaged azimuthally and over time to produce the intensity as a function of scattering vector $I \left(q\right)$. Finally, this curve was normalized by the incident flux, detector efficiency and area, and the solid angle spanned by the detector. In the results presented below, we report the normalized intensity curve
\begin{equation}
I_N\left(q\right) = d \, \Tr \, \phi \, V_{part} \, \left( \Delta \rho \right)^2 \, P\left(q\right) \, S\left(q\right) \equiv A \, P\left(q\right) \, S\left(q\right),
\label{myinteqn}
\end{equation}
where $d$ is the cell thickness, $\Tr$ is the transmission coefficient, $V_{part}$ is the average particle volume, and $\Delta \rho$ is the electron density difference between silica and water or ice. The coefficients are grouped together into the amplitude $A$. We did not normalize by the sample thickness or transmission because, due to the pressure difference a slight curvature was present, so the sample thickness was not known precisely at each sampling position. In what follows, we will refer to the normalized intensity as simply $I \left(q\right)$. A more comprehensive background to x-ray scattering can be found in the appendix and the references therein.

% **** DIRECT IMAGING **** %

\section{Direct Imaging}

Before delving into the SAXS results, we present direct images of freezing and frozen colloidal suspensions under conditions similar to those used in the x-ray scattering experiments. These images provide a reference for interpreting the features in the SAXS intensity curves.

We used the same sample cell and type of colloidal solutions in the direct imaging experiments as in the SAXS experiments. In addition to colloidal samples, we also observed samples without particles that were simply pure deionized water. The cell was situated between the light source and the camera, thus the samples were viewed in transmission. Images were focused onto a CCD detector (Unibrain Fire-i) with a $4$x microscope objective lens resulting in an image scale of about $6 \, \mu$m per pixel.

We froze the samples by lowering the temperature of the cold finger either directly with the TECs or with liquid nitrogen. The samples typically froze at temperatures between $-6^{\circ}$C and $-25^{\circ}$C, though the freezing temperatures of individual samples had a high degree of uncertainty (up to $\pm 2^{\circ}$C). In all cases, the water was supercooled when ice nucleated, resulting in two stages of ice growth: a rapid stage I with a cellular or dendritic morphology, and a slower stage II with an apparently planar morphology.

\begin{figure}[tp]
\begin{center}
\includegraphics[]{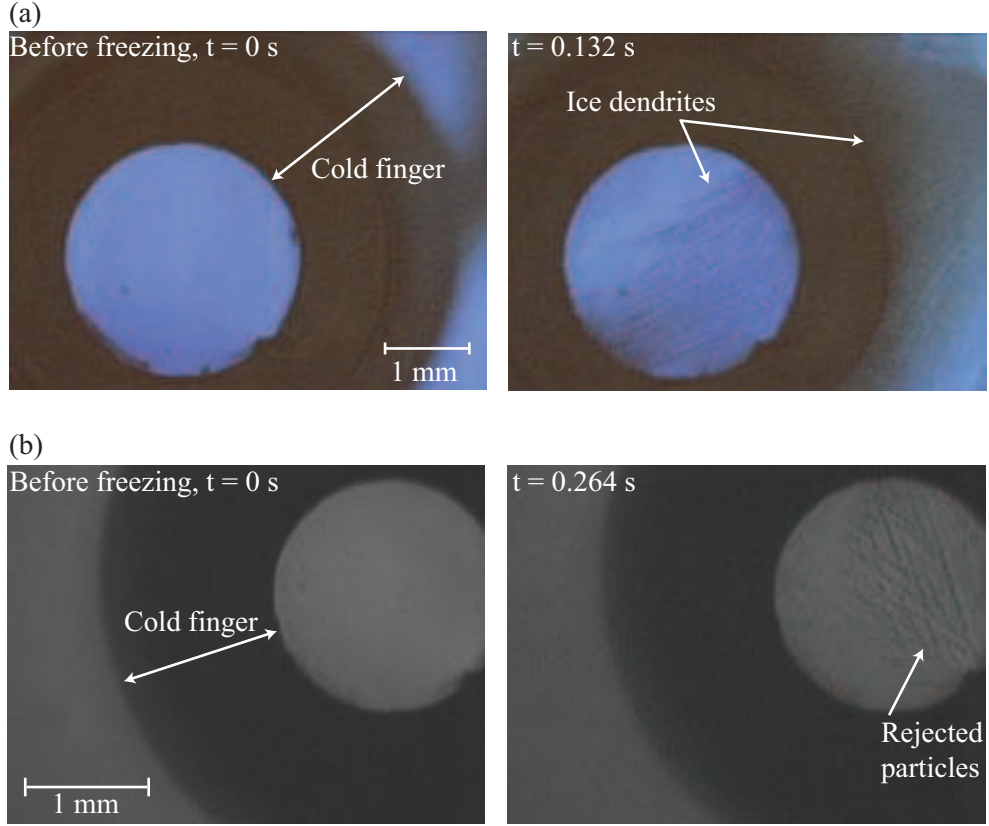}
\caption[Examples of stage I ice growth]{These images show two sets of before (time $t = 0$ s) and after ($t>0$ s) snapshots from movies of stage I ice growth. The images in (a) show pure water, whereas those in (b) show a colloidal solution of silica spheres as described above, but with particle radius $142$ nm. We note that we did not observe any significant differences in the direct imaging experiments between the behavior of solutions of these larger particles and solutions of the smaller particles (as used in the x-ray scattering). Ice dendrites are visible in both sets: dark in (a) and lighter areas between dark regions of concentrated particles in (b).}
\label{stageI}
\end{center}
\end{figure}

During stage I, the low temperature of the sample caused rapid solidification and ice growth into a solution below the bulk melting temperature $T_m$, leading to an unstable solidification front and a cellular or dendritic ice growth morphology \cite{mullins1963,shibkov2003,peppin2007,deville2009}. Figure \ref{stageI} shows two sets of images obtained from movies of the sample freezing that illustrate this stage of ice growth in pure water (a) and a colloidal solution (b). The ice growth is cellular or dendritic with a linear pattern of alternating dark and light lines visible inside the cold finger in both samples. Because the entire field of view often froze in the time span of only a few frames (at frame rates of $7.5$ or $15$ fps), estimates of the freezing rates have large uncertainty. However, the values mostly fall between $10$ mm$/$s and $40$ mm$/$s, which agree fairly well with the morphology diagram for pure water in reference \cite{shibkov2003}.

\begin{figure}[tp,floatfix]
\begin{center}
\includegraphics[]{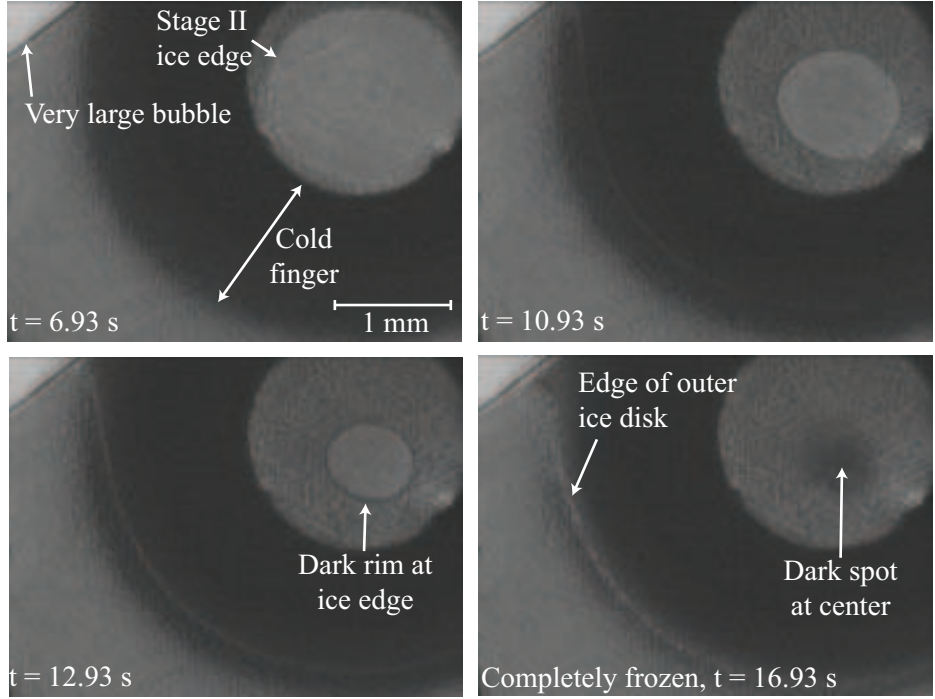}
\caption[Example of stage II ice growth]{These images show a sequence of snapshots of stage II solidification for a sample of $142$ nm particles. In areas where stage II ice has formed the sample appears darker. The stage II ice edge is marked by a dark rim of particles being pushed ahead of the ice, which form a dark spot at the center upon complete solidification.}
\label{stageII}
\end{center}
\end{figure}

Stage I freezing ended when the entire sample had been warmed to $T_m$ through release of latent heat of solidification. After this time, further freezing required further removal of heat from the sample, which was effected by the TECs. We then observed an apparently planar ice front growing radially inwards and outwards from the cold finger, freezing any water that remained after stage I. Figure \ref{stageII} contains a sequence of images showing this stage II ice growth, during which the ice edge moves radially inwards at a constant rate of $0.085$ mm$/$s. Because the solidification rate during stage II is slower, measurements are much more accurate. All freezing rates are nearly constant throughout stage II ice growth and vary between about $0.1$ mm$/$s and $1$ mm$/$s among the samples. Although the stage II ice front appeared to be stable and planar, in fact it may have been unstable, just with a wavelength below the resolution of our imaging setup. Previous work \cite{waschkies2009} has shown that for solidification rates in the range of our experiments, the wavelength of the instability drops below $10 \, \mu$m, which we would not be able to resolve.

\begin{figure}[tp]
\begin{center}
\includegraphics[]{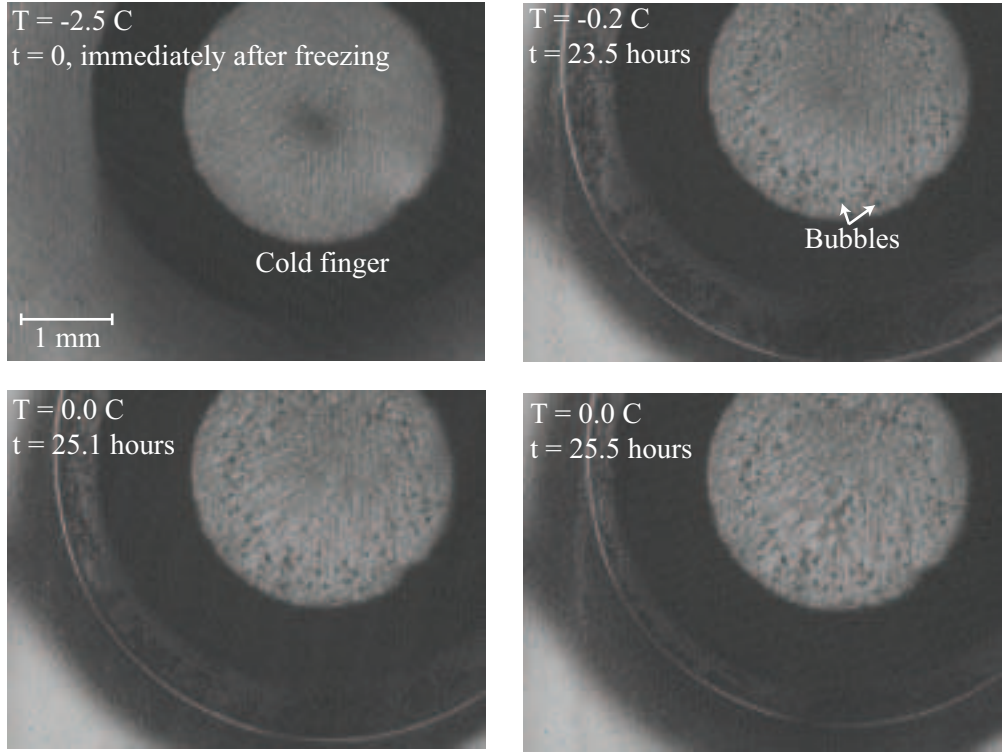}
\caption[Example of pattern evolution]{These images show a sequence of snapshots of a frozen solution of $32$ nm particles over time with the sample ages and temperatures indicated. The small dark spots near the cold finger are air bubbles. Because the water was not degassed before freezing, air gradually exolves from the ice.}
\label{evolution}
\end{center}
\end{figure}

During stage II, the linear pattern of light and dark stripes formed during stage I disappears from pure water samples, whereas it persists in colloidal samples. This pattern is evident both in Fig. \ref{stageII} and in the first image of the sequence in Fig. \ref{evolution}. Because the samples were viewed in transmission, areas of high particle density should appear dark whereas areas of low particle density should appear light. Therefore, we interpret the light and dark stripes present in colloidal samples as a pattern of high and low particle density imposed by the ice during freezing. In the pure water samples, they are simply an optical effect due to the edges of the dendrites, which disappear once the stage II ice growth has solidified all water remaining between the dendrites.

These observations indicate that the particles were rejected to the inter-dendrite regions during stage I and then engulfed by the ice during stage II. The critical freezing rate above which a single particle at a planar ice interface will be engulfed by the growing solid depends upon the particle size and the specific intermolecular interactions between the particle and the solid \cite{rempel1999}. Using the magnitude of the interaction between glass particles and ice determined by reference \cite{peppin2009}, we find that the particles should have been rejected from the growing ice during both stages \cite{mythesis}. Indeed, during stage I the particles were rejected into the inter-dendrite regions, and during stage II some of the particles not in the inter-dendrite regions were evidently rejected as dark patches appeared at the center of the cold finger in some samples. However, the particles rejected to the inter-dendrite regions were engulfed by stage II ice growth, perhaps by trapping between the dendrites \cite{wilde2000, deville2009} or possibly due to their inclusion in large particle aggregates \cite{mythesis}. Such effects are not included in the model of reference \cite{rempel1999}, thus highlighting the need for ongoing work. In summary, the process of freezing supercooled colloidal suspensions in our experimental setup results in a linear pattern of high- and low-particle-density regions due to the unstable freezing morphology.

The width of the stripes was typically tens of $\mu$m; for the sample shown in Fig. \ref{evolution}, initially the light (low density) regions were on average $17 \, \mu$m across and the dark (high density) regions were on average $28 \, \mu$m across. However, these patterns and the widths of the respective regions changed as the temperature increased and the samples aged. We observed the evolution of the samples at different temperatures between $-2^{\circ}$C and $0^{\circ}$C over time scales ranging from several hours up to one week. In general, the light areas became more rounded, and the linear dark features tended to merge with each other, their edges becoming simultaneously more sharply defined. Figure \ref{evolution} contains a sequence of images illustrating this evolution. Our direct observations and dynamic x-ray scattering suggest that this evolution is driven by grain boundary motion due to coarsening of the polycrystalline ice in the samples \cite{spannuthdspreprint}.

Upon melting, we observed that many dark objects up to $100 \, \mu$m in size sedimented out of the solution. Presumably, these were aggregates of individual particles bound together during the freezing or subsequent evolution processes (images provided in reference \cite{mythesis}).

Finally, we note that all of the observations described above occurred in a qualitatively similar manner despite differences in the initial freezing temperature and how the temperature changed over time after freezing. While lower freezing temperatures resulted in faster solidification velocities \cite{mythesis}, and thus presumably differences in particle incorporation as well as the micrometer-scale structure \cite{deville2009b}, our direct imaging experiments had insufficient resolution to quantify these variations. However, as we will describe in the next section, such differences do not significantly affect the particle-scale structure, particularly the inter-particle spacing.

In summary, the direct imaging experiments provide general information about the freezing process in our system. After deep supercooling, the initial stage of solidification is unstable with particles being rejected to the regions between ice dendrites to form a linear pattern of high and low particle density. During the second stage of solidification, this pattern is locked in as an apparently planar ice front grows across the cell. As the frozen samples evolve, the high-particle-density regions rearrange due to the motion of grain boundaries from ice crystal coarsening. These observations serve as a framework for understanding the results of the x-ray scattering experiments, which provide quantitative information about the particle-scale structure in these macroscopic features.

% **** SAXS RESULTS **** %

\section{SAXS Results}

\begin{figure}[tp]
\begin{center}
\includegraphics[]{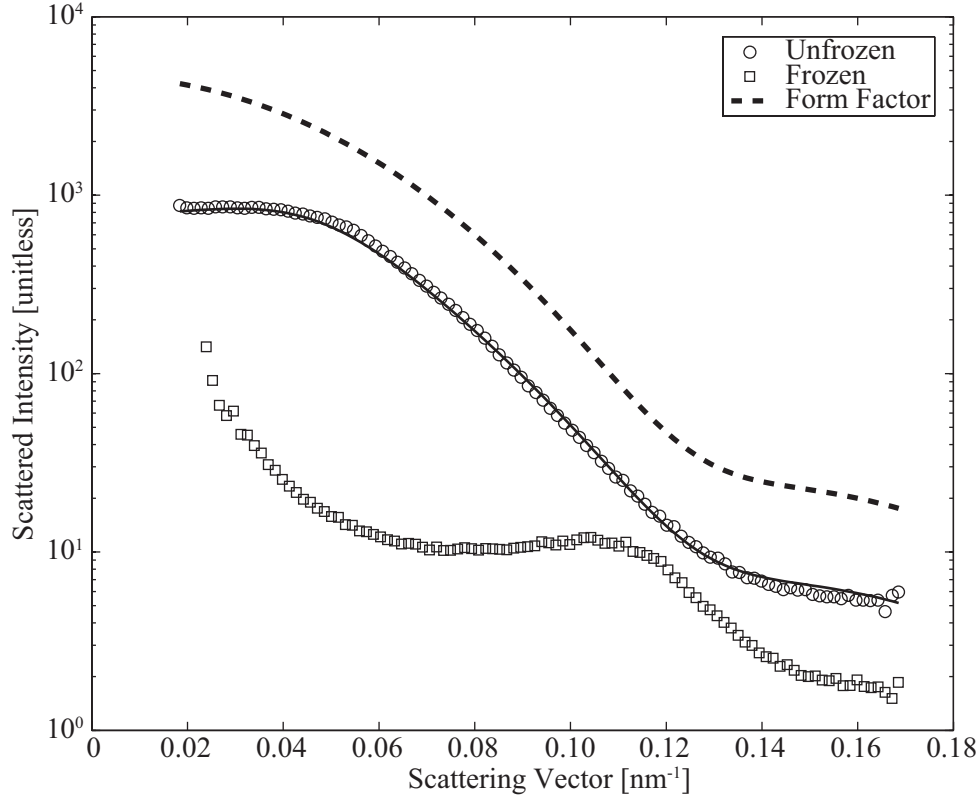} 
\caption[SAXS intensity versus scattering vector]{SAXS intensity versus scattering vector taken at the center position from sample $1$ before being frozen (circles) and at $T = -2.00^{\circ}$C when frozen (squares). The solid curve represents the fit of the unfrozen data to a polydisperse sphere form factor and monodisperse hard sphere structure factor with $R = 32.4$ nm, $z = 31$, $A = 294$, $R_{HS} = 53.0$ nm, and $\phi_{HS} = 0.073$ as described in the text. For comparison, the dotted line shows only the form factor with the same parameters as above, but an arbitrary amplitude. For clarity the unfrozen data have been offset from the frozen data by multiplication with a constant coefficient.}
\label{SAXSfigure}
\end{center}
\end{figure}

The primary result of SAXS is the scattered intensity $I\left(q\right)$. Figure \ref{SAXSfigure} shows typical examples of $I\left(q\right)$ for a sample before being frozen (circles) and when frozen (squares). The unfrozen data decrease smoothly as $q$ increases, whereas the frozen data have two features: a peak at high $q$ and an upturn at low $q$. For all temperatures at which the sample was frozen, the intensity maintained the same general form with these two features, though the position and width of the features changed. Upon melting, the scattered intensity reverted to the unfrozen form observed before the samples were frozen, though the details of the shape had changed. These data reflect the structural properties of the samples such as the particle size, shape, and inter-particle spacing. By fitting the intensities to a theoretical model (unfrozen data) and an empirical function (frozen data), we were able to quantify these structural properties and monitor how they evolved as the sample temperature was increased and the samples aged.

\subsection{Unfrozen Intensity}

For the unfrozen intensities, we obtained the particle radius, polydispersity, and volume fraction by fitting the data to a function of the form $I \left(q\right) = A \, P\left(q\right) \, S\left(q\right)$ from equation \ref{myinteqn}, where $A$ is a $q$-independent coefficient signifying the amplitude of the scattering, $P\left(q\right)$ is the particle form factor, and $S\left(q\right)$ is the structure factor. Though the particles are not perfectly spherical, we used a standard form factor for polydisperse spheres \cite{aragon1976} that depends on the average particle radius $R$ and the polydispersity parameter $z$. This form factor is based on a Schulz-Zimm distribution of individual particle radii $R_p$ in which $z$ describes the width of the distribution. In this case, the mean square deviation is given by $\overline{R_p^2} / \left(z+1\right)$ where $R = \overline{R_p}$ \cite{aragon1976}. For the structure factor we used a function for monodisperse spheres of radius $R_{HS}$ at volume fraction $\phi_{HS}$ interacting via a hard sphere potential \cite{lumma2000pre}. For each unfrozen data set, the fitting was performed using an iterative grid search method to find the parameters that minimized the mean squared residual. Due to the large range of intensity values, the logarithm of the data was used to determine the residuals. We also visually inspected each fit to ensure quality. The solid curve in Fig. \ref{SAXSfigure} shows this fit to a typical data set.

Altogether there are five parameters in the fitting equations: the average particle radius $R$, the polydispersity $z$, the hard sphere radius $R_{HS}$, the hard sphere volume fraction $\phi_{HS}$, and the amplitude constant $A$. Across all samples and positions, the average particle radius was $32 \pm 1$ nm and the polydispersity $29 \pm 3$ both before freezing and after melting, for a spread of about $18 \%$ around the average radius (as given above in section \ref{materials}). The hard sphere radius was typically $53 \pm 1$ nm with the change between the pre-freezing and after-melting values negligible compared with uncertainty in the fitting. In contrast, the hard sphere volume fraction was typically between $0.07$ and $0.09$ before freezing, but usually dropped to between $0.03$ and $0.07$ after melting. We do not discuss $A$ here because without measuring the absolute scattered intensity, changes in this parameter cannot be interpreted unambiguously. Thus, most parameters did not change significantly from before freezing to after melting (to within the uncertainty in the fit), except that the hard sphere particle volume fraction decreased by a factor of $2$ or more. 

The average particle radius and the polydispersity reflect the actual physical extent of the particles. Thus, the near constancy of these parameters indicates that the physical size of individual particles and distribution of those sizes did not change during freezing or subsequent evolution. The hard sphere radius, on the other hand, represents the effective radius of the particles in their interactions with each other (assuming they interact according to a hard sphere potential). Because $R_{HS}$ is larger than $R$, the particles apparently \textit{behaved} as if they were larger than their physical dimension. As a result, $\phi_{HS}$ overestimates the actual particle volume fraction $\phi$. The two volume fractions can be related by $\phi = \phi_{HS} \left(R / R_{HS}\right)^3$. This gives initial actual volume fractions of about $0.015$ -- $0.02$ and final actual volume fractions between $0.007$ and $0.015$. As $R_{HS}$ did not change significantly throughout the experiment, the decrease in $\phi_{HS}$ represents a real decrease in the bulk particle concentration from before freezing to after melting.

\subsection{Frozen Intensity}

\begin{figure}[tp]
\begin{center}
\includegraphics[]{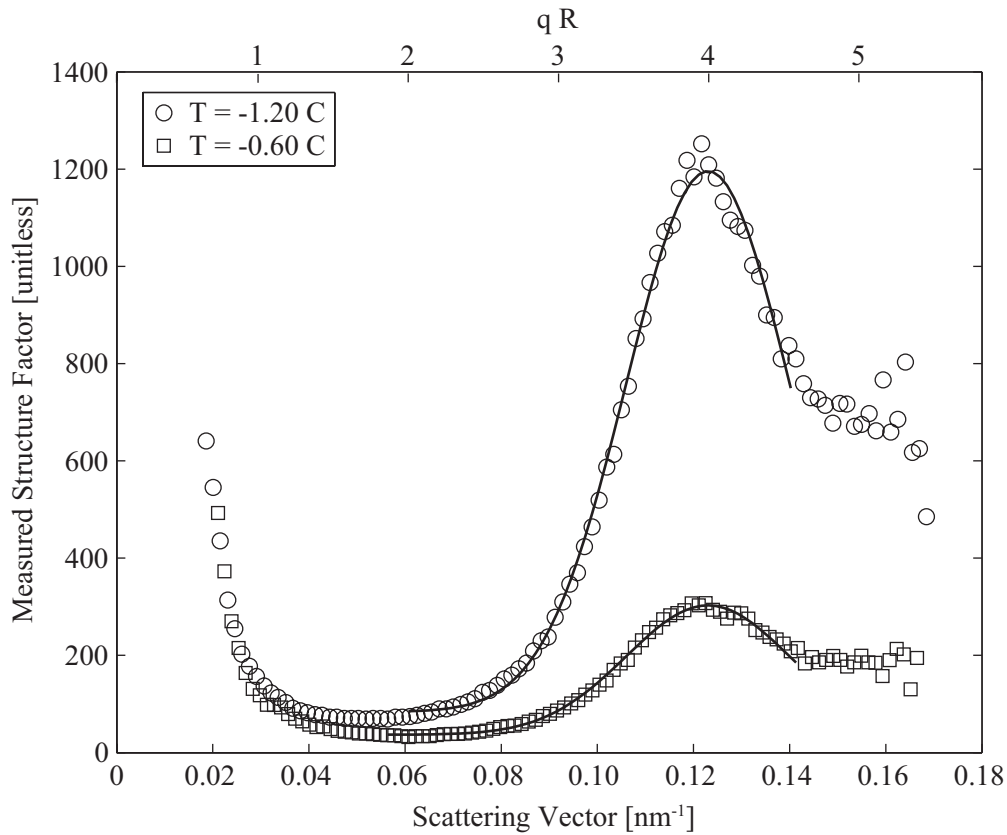} 
\caption[Measured structure factors versus scattering vector]{Measured structure factors versus scattering vector taken at the left edge position from sample $4$ at $T = -1.20^{\circ}$C (circles) and $T = -0.60^{\circ}$C (squares). Solid curves represent the Gaussian fits of the main peak as described in the text. Along the top of the plot, the horizontal axis is labeled in units of $q R$.}
\label{measstructfact}
\end{center}
\end{figure}

For the frozen data, we isolated the structure factor by dividing the intensities by the form factor used for the unfrozen solutions with $R = 32$ nm and $z = 29$. Because $I \left(q\right) = A \, P\left(q\right) \, S\left(q\right)$, dividing by $P\left(q\right)$ leaves a measured structure factor $S_m\left(q\right) = A \, S\left(q\right)$. Examples of $S_m\left(q\right)$ are shown in Fig. \ref{measstructfact}. Like the full intensity profile, the measured structure factor has a clear peak at high $q$-vectors and an upturn at low $q$-vectors. Whereas the upturn is more prominent in the full $I\left(q\right)$ because it is enhanced by the large values of $P\left(q\right)$ at low $q$ (see dashed line in Fig. \ref{SAXSfigure}), conversely the peak is enhanced in $S_m\left(q\right)$. The upturn in the structure factor at low scattering vectors represents structure on length scales larger than several times the particle radius. The peak at higher scattering vectors reflects structure on the single particle length scale, giving information about the particles' nearest neighbors. 

In order to obtain information about the particle packing, we attempted to fit $S_m\left(q\right)$ with a variety of common structure factors with $A$ as a free parameter. We were unable to obtain acceptable fits with structure factors derived from a monodisperse hard sphere potential \cite{lumma2000pre}, polydisperse hard sphere potential \cite{vrij1979,vanbeurten1981}, sticky hard sphere (square well) potential \cite{dawson2000, pontoni2003}, or Coulomb repulsion \cite{hayter1981}. In part, the failure of the structure factor models resulted from their inability to reproduce the upturn at low $q$. Therefore, we also investigated fitting only the high $q$-vector peak, yet we were still unable to obtain acceptable fits with any of the hard sphere models. In a further attempt to fit the low $q$ upturn, we modified the form factor by including a fractal cluster term \cite{pedersen2002} or a $q^{-4}$ dependence \cite{pontoni2003}, but neither improved the fits.

Instead, we fit the main, high $q$-vector peak with a Gaussian function given by
\begin{equation}
I\left(q\right) = \delta + \alpha \exp \left[ - \left( q-q_{peak} \right) ^2 / \sigma^2 \right],
\end{equation}
where $\delta$ is the $q$-independent offset, $\alpha$ is the $q$-independent peak height, $q_{peak}$ is the peak location, and $\sigma$ controls the peak width. In order to obtain reliable fits, we only used data between chosen low- and high-$q$-vector cutoffs. The low-$q$-vector cutoff was that scattering vector at which the measured structure factor reached its minimum value. The high-$q$-vector cutoff was defined as $q = 0.14$ nm$^{-1}$. We fit the plain values of $S_m\left(q\right)$ rather than their logarithm to emphasize fitting of the peak. The fitting was performed using an iterative grid search method to minimize residuals. As with the unfrozen data, we visually inspected the resulting fits to ensure good quality. The solid curves in Fig. \ref{measstructfact} illustrate these fits.

\begin{figure}[tp]
\begin{center}
\includegraphics[]{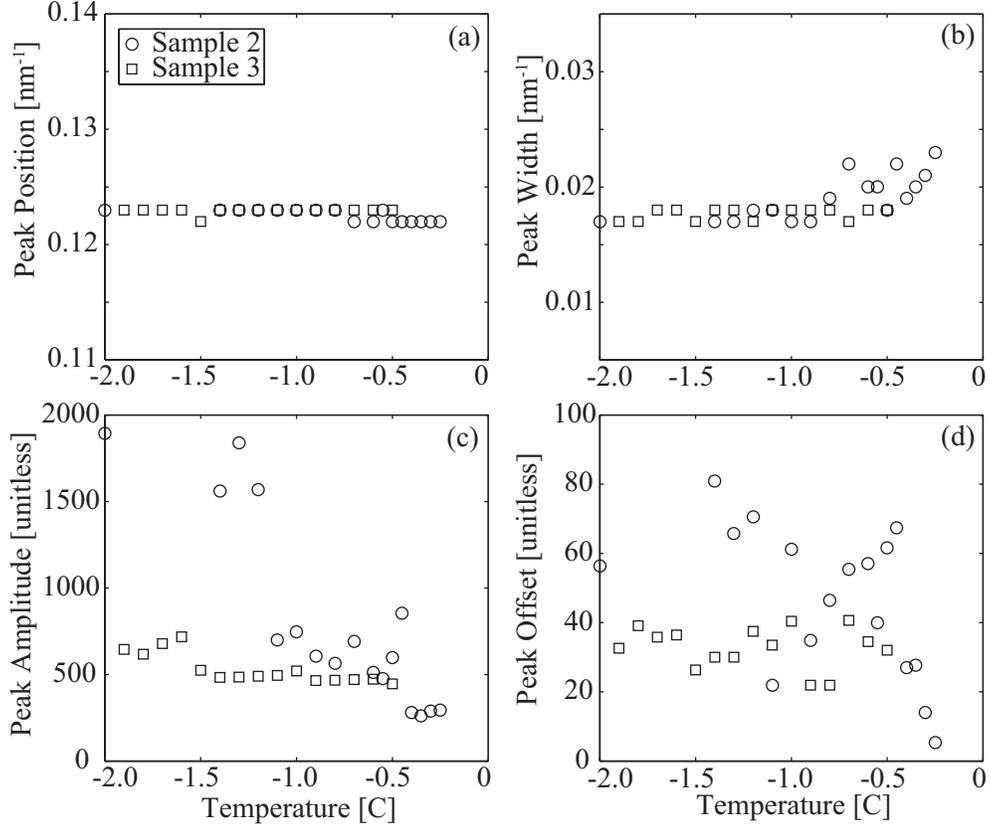} 
\caption[Peak fit params]{Peak fit parameters versus temperature from the center position of two different samples (circles, sample $1$; squares, sample $2$). The peak position is in (a), the peak width is in (b), the peak amplitude is in (c), and the peak offset is in (d).}
\label{peakparams}
\end{center}
\end{figure}

We performed this Gaussian fitting on all data sets for which the samples were frozen and examined the resulting fit parameters as a function of temperature (examples are shown in Fig. \ref{peakparams}). In general, all data exhibit similar trends, though there is some variation among positions within a given sample and among different samples. This variation is produced by the inherently stochastic nature of the ice nucleation process, the unstable ice growth morphology, and the process of ice crystal coarsening in the polycrystalline ice. These processes lead to spatial variations in the total number of particles contained within the scattering volume and differences in how this number changes with time and temperature. Such variation in particle number primarily affects the fitted values of $\alpha$ and $\delta$.

We find that the peak position and peak width, which represent the predominant inter-particle spacing (nearest neighbor distance) and the distribution of inter-particle distances, are fairly constant at $q_{peak} \approx 0.123$ nm$^{-1}$ ($q_{peak} R \approx 3.94$) and $\sigma \approx 0.017$ nm$^{-1}$, respectively, though the peak width appears to increase slightly in some cases. This indicates that the average inter-particle distance remained fairly constant while the samples were frozen. The increasing peak width indicates that the distribution of inter-particle distances widened slightly. Surprisingly, variations in ice nucleation temperature, and hence freezing rate, do not translate into variations in particle spacing. The peak amplitude shows a very clear decreasing trend as the temperature increases, while the offset does not exhibit a clear trend. The offset is simply related to the overall amount of scattering, which we expect to change between data sets as particles move into or out of the scattering volume. The decreasing peak amplitude indicates a decrease in the number of nearest neighbors. When combined with the increasing peak width, this suggests an increase in heterogenity of particle spacing as the temperature increased and the samples aged.

Altogether, SAXS reveals that the scattered intensity, and hence the sample structure, changed very dramatically when the samples froze and continued to evolve while the samples were frozen. The changes in $I\left(q\right)$ from before freezing to after melting indicate that the volume fraction of particles within the bulk solution decreased significantly. While the samples were frozen, the steady position of the high $q$-vector peak shows that the average nearest neighbor distance remained fairly constant. The slight widening of the peak and the decrease in its amplitude suggest that the distribution of inter-particle distances widened and became more heterogeneous. In the next section, we interpret these results within the context of the direct images obtained from our laboratory experiments and with respect to established models for the structure factor.

% **** DISCUSSION **** %

\section{Discussion}

These SAXS results provide quantitative information about the structures observed in the direct images, and conversely, the direct imaging experiments provide a qualitative framework for interpreting the SAXS results. In particular, direct observation revealed a linear pattern of high and low particle density that formed during the dendritic freezing of the colloidal solutions at high levels of supercooling. This pattern subsequently evolved as the temperature increased and the samples aged, with regions of high density joining together and regions of low density enlarging. Furthermore, we observed particle aggregates sedimenting out of solution as the samples melted. Each of these observations can be identified with and quantified by features in the SAXS results.

For the following discussion it is important to note the relative size of the x-ray beam as compared with the pixel size in the direct images in order to maintain the proper perspective on the structures probed by the x-ray scattering. As the size of a single pixel in the direct imaging setup was approximately $6 \, \mu$m square, the entire x-ray beam (approximately $20 \, \mu$m square) covered roughly an equivalent area of $9$ pixels ($3$ pixels by $3$ pixels) in the direct images. Therefore, the scattering volume probed by the x-rays, and consequently the structures inferred from the scattering data, are comparable to the smallest details that could be observed in the direct imaging experiments.

We first discuss the scattering data from unfrozen solutions. These data were fit to a model based upon polydisperse spherical particles that interacted as if they were monodisperse hard spheres. In fact, before being frozen the colloids likely interacted according to DLVO theory \cite{israelachvilibook} because the attractive van der Waal's interaction at short range was counteracted by the long-range repulsive electrostatic interaction arising from surface-induced ionization. Generally, silica colloids have silanol (SiOH) surface groups which ionize in solution to form negatively charged SiO$^{-}$ groups that give the particles an overall negative charge and hence stabilize the solution \cite{robertsbook}.The H$^+$ ions that dissociated from the surface mix with any other ions in the water and form a diffuse layer of higher ion concentration surrounding each of the particles with characteristic thickness given by the Debye length \cite{israelachvilibook}.

Several studies of colloids have found differences between the physical particle radius as measured by electron microscopy and the actual radius or effective hard sphere radius measured by static or dynamic light scattering \cite{philipse1988, philipse1989, imhof1994, bryant2003, schope2007}, and indeed charge stabilized colloids have been found to behave as effective hard spheres \cite{nagele1997}. In our system, the SAXS data show that when not frozen the particles could be treated as hard spheres with an effective hard sphere radius somewhat greater than the actual particle radius. Presumably, the effective hard sphere radius is larger than the actual radius due to the cloud of ions surrounding the particles, but an exact relationship between the effective hard sphere radius and the Debye length is not known. Thus, we attribute the difference between $R$ and $R_{HS}$ to the dissolved ionic species remaining in the solutions.

Next, we consider the change in $\phi$ between the SAXS measurements made before freezing the samples and after melting. The particle volume fraction obtained after adjustment from the fitted $\phi_{HS}$ tended to drop from about $0.02$ before freezing to around $0.01$ after melting. This decrease indicates that after being frozen and melted, the bulk solution contained less than half as many particles as it contained before being frozen. The missing particles presumably sedimented as aggregates, as we observed in the direct imaging experiments. Because the particles used in our experiments have a very small Peclet number (about $10^{-5}$), Brownian motion is sufficient to keep individual particles suspended almost indefinitely. However, the increased mass of particle aggregates could cause sedimentation on experimentally relevant time scales \cite{mythesis}. Therefore, the SAXS results from the unfrozen samples imply that about half of the particles originally in the solution ended up in long-lived aggregates and subsequently sedimented upon melting of the ice. The aggregates most likely formed in the high particle density regions created by rejection of particles from the ice dendrites. This is supported by the interpretation of the SAXS data collected while the samples were frozen.

The scattered intensity from frozen samples had two primary features associated with structure possessing two distinct primary length scales. The high scattering vector peak corresponds to the inter-particle spacing of colloids within the high density regions and the low scattering vector upturn is related to the size of the high density domains. If we could extend our measurements to lower scattering vectors, we would expect to find that the upturn is in fact a peak and its position would give the size of the high density domains (as in references \cite{stradner2004,lattuada2004,sciortino2004}) or the spacing between them. In the present experiments, the minimum $q$ gives a lower bound for the size of these features: $2 \pi / q_{min} = 2 \pi / 0.02$ nm$^{-1}$ $= 314$ nm, or approximately $10$ times the particle radius. Further measurements at lower scattering vectors would also help clarify the medium-scale structure of the high density regions, i.e., the arrangement of particles on length scales greater than that of a single particle, but still within a single high density region. Although the failure of the fractal cluster model to fit the low-$q$ upturn in our data {\em suggests} that structure at this scale is not fractal, there are insufficient data to rule this out or to advance other possibilities.

On the other hand, the high scattering vector feature provides more reliable information because the full peak falls within our accessible $q$-range. This peak reflects how the particles packed as they were rejected during freezing. We can rule out a crystalline arrangement of the particles because the peak is too broad. We did not expect that the particles in the present experiments would pack this way due to their large polydispersity, which is known to inhibit colloidal crystallization \cite{henderson1996,auer2001,martin2003}. In addition, colloidal crystallization is an equilibrium process requiring some amount of time to proceed. Although an ordered particle packing has been observed in at least one directional solidification experiment \cite{kim2009}, the densification of the particles upon rejection during freezing in our samples was most likely too rapid to permit this process \cite{auer2001b}. Therefore, the particles in the high density regions packed in a predominantly amorphous or random arrangement.

Particles in an amorphous packing, like particles in a colloidal crystal, are characterized by an average inter-particle distance though the variation around this average distance is greater in amorphous packings than in crystalline ones. The position of the SAXS peak is approximately related to this distance by $2 \pi / q_{peak}$, which gives an interparticle distance on the order of the particle diameter for our data. Therefore, we conclude that the particles in the high density regions were generally in contact with their nearest neighbors. However, this is insufficient to determine the particle volume fraction. Knowing that on average particles were in contact with their nearest neighbors offers no information about how many nearest neighbors an average particle contacts, which is related to $\phi$.

Typically the volume fraction is quantified through the model for $S\left(q\right)$. However, our measured structure factors did not conform to structure factors based on common particle pair potentials. Therefore, we estimate the volume fraction by analogy with another experiment on dense, polydisperse colloidal suspensions. Pham and colleagues \cite{pham2004} suggested that a shift of the peak position in their scattering data from $q R \approx 3.8$ to $q R \approx 4.0$ corresponded to a change in the local particle volume fraction from $0.60$ to $0.69$, the random close packing limit for their system. The enhancement above the often-quoted random close packing value of $0.64$ was attributed to particle polydispersity. We note that they did not compare their data with any models. Based on their empirical relation and the location of our peak at $q R \approx 3.94$, we estimate that the particles in the high density regions had a volume fraction near $0.66$. This is similar to the predictions from simulations for spheres of similar polydispersity, which range from $0.66$ to $0.68$ \cite{schaertl1994,farr2009}. Thus, the particles in the high density regions were likely at their close-packing limit.

With this knowledge, we can explain the formation of the observed particle aggregates. Before the particles can aggregate though, they must be brought into contact. In order to bring the particles into contact, the ice must have exerted a force on the particles sufficient to overcome the repulsion between two particles resulting from the surface charges and double layer. From frost heaving of soils the maximum overpressure at which heaving stops has been measured at about $11$ atm per $^{\circ}$C of cooling below $T_m$ \cite{dash2006}. For the present experiments where freezing occurred below $-20^{\circ}$C the pressure on the particles may have been larger than $200$ atm, which is much larger than the expected electrostatic repulsion. Thus, it is reasonable that the ice should be able to overcome the repulsive force between the particles and push them into close contact.

Once this repulsion was overcome and the particles were forced into contact by the ice dendrites, the attractive van der Waal's force should have dominated the interaction, allowing the particles to form aggregates. Using the Hamaker constant for fused quartz and a separation of $0.25$ nm (the approximate size of a water molecule), the attraction potential between two particles is estimated to be $-9.4 \times 10^{-20}$ J, or about $25 k_B T$ at $T = 0^{\circ}$C, where $k_B$ is Boltzmann's constant \cite{israelachvilibook}. This is sufficient to maintain the aggregates' integrity well above the melting temperature. Alternatively, once the particles were forced into contact they may have fused together chemically or physically due to damage near the inter-particle contacts, which may also have been responsible for the failure of standard structure factors to fit our data. In either case, this suggests that the aggregates were the direct result of particle rejection to the inter-dendrite regions during freezing. Combined with the SAXS results from the solutions when unfrozen, we can then estimate that at least half of the particles in the solution ended up in these high density regions. However, this is a lower bound, and in fact nearly all of the particles may have been caught between the dendrites, but some were individually engulfed by the ice or subsequently stripped from the aggregates by dynamic processes \cite{spannuthdspreprint}.

The strong forcing of the particles by the growing ice during freezing may also be responsible for the failure of standard structure factors to model our scattering data. This failure implies that the particle configuration within the high-particle-density regions was different from those that occur in high density colloidal fluids (c.f. reference \cite{lumma2000pre}), glasses (c.f. reference \cite{lu2008}), or gels (c.f. reference \cite{shah2003}) even though the underlying particle interactions are similar (long-range repulsive or hard sphere interactions and short-range attraction). In particular, the peak in our data generally occurred at higher scattering vectors and was broader and taller than the peak predicted by any of the standard models. This means that our samples tended to be more heterogeneous with respect to the inter-particle spacing and have a greater number of nearest neighbors than expected from these models. Such differences may be related to the forcing present during freezing.

Although much work has been directed at studying the influence of shear flow on structure in various colloidal materials (e.g., references \cite{haw1998,varadan2001,vermant2005}), relatively little work has been done on other types of external forcing that are more comparable to what the particles experience during freezing. One example, though, is the experiment of Kurita and Weeks \cite{kurita2010}. They examined a layer of randomly close packed, sedimented particles using confocal microscopy and calculated a structure factor from the real-space positions of the particles. Their system had an overall volume fraction of about $0.646$ with small, locally ordered regions having $\phi$ up to about $0.68$ and they found that the resulting $S\left(q\right)$ had a primary peak near $qR \approx 3.93$ (versus $3.94$ for our system). This peak position is higher than expected for hard spheres at these volume fractions, but they did not attempt to fit their structure factor to any models, so differences in the shape are not known. Sedimentation involves a gradual increase in the particle density and compression of the colloidal fluid, similar to what happens during freezing when the growing solid continually squeezes the particles into the shrinking volume of unfrozen liquid. Therefore, we might expect that our samples had a structure with characteristics similar to that of the sedimented layer: very high densities with some local variability. However, the rate of compression during freezing is much higher than in sedimentation, which could lead to more variability in the packing and a broader structure factor peak. In addition, the morphology of ice growth and the kinetics of particle segregation could also lead to more heterogeneity in the packing.

Altogether, our observations have several implications for our understanding of directional solidification of colloidal suspensions. First, models based on the purely statistical mechanical behavior of colloidal \textit{solutions}, while a reasonable and necessary starting point, are likely not adequate to completely explain the phenomena observed during solidification. That is, the densification that occurs as the solutions freeze is not analogous to that resulting from simply increasing the density of particles in a colloidal fluid. Second, the results add experimental evidence to the common assumption that the colloidal particles close pack upon rejection from the solidification front and suggest that the packing achieved may be the densest possible amorphous packing that can be produced given the particles' distribution of sizes. Furthermore, the high particle density regions appear to be compact (i.e., not fractal) on the scale of several particle diameters. Finally, the observation of particle aggregates whose attractive van der Waal's interaction is sufficient to maintain their integrity after melting suggests the possibility of creating macroscopic freeze-cast materials without the need for special binding or sintering techniques.

In addition, our results may be useful in understanding other systems involving driven, high concentration colloidal suspensions. They suggest that the arrangement of particles at the smallest scales may not conform to predictions based solely on the interparticle interactions. Such differences could potentially influence the flow properties of the material or dynamic behavior of the particles, which are of interest scientifically and for engineering applications. Further study of the structure factor could incorporate hydrodynamic interactions between the particles and the effects of the driving force (such as repulsion from the ice), as has been done for the well characterized shear flow geometry \cite{vermant2005}. Extending such work to more complicated forcing configurations and flow geometries is important because these types of situations are often encountered in practical applications. Overall, solidified colloidal suspensions are a promising system in which to study the effects of external driving on particle arrangement because the particle-scale structure is effectively ``frozen in'' both by the constraint of the surrounding ice and the strong van der Waal's attraction between the particles.

% **** CONCLUSION **** %

\section{Conclusion}

We have presented a combined small angle x-ray scattering and direct imaging study of frozen colloidal suspensions. Our results highlight the utility of these methods for studying the structure of such materials and suggest routes for future investigation. By consulting the images acquired directly in laboratory freezing experiments, we identified the main peak exhibited by the scattered x-ray intensity from frozen solutions as resulting from the close packing of particles in high-particle-density regions formed between ice dendrites. The enhanced intensity at low scattering vectors we attributed to the size of the high density regions. In addition, the close packing of the particles produced by freezing allowed the short-range attractive inter-particle interaction to dominate thereby creating long-lived particle aggregates. However, we found that the structure observed in our samples when they were frozen could not be described by any of the standard inter-particle potentials even though the unfrozen solutions were well-described by a hard sphere interaction with an effective hard sphere radius. This implies that the process of freezing produces atypical arrangements of the colloidal particles.

Further work could help clarify some of the issues encountered and expand upon the present conclusions. Importantly, by altering the solidification conditions, more controlled freezing could be attained and particle structure (including volume fraction) ahead of a solidification front (planar or dendritic/cellular) could be studied. By using a linear solidification geometry, as opposed to the radial geometry in the present experiments, we could better connect our observations with the original work on directional solidifiction of particle suspensions \cite{korber1985,lipp1990}. Examining samples with a variety of higher and lower initial volume fractions would help determine the robustness of the close packed arrangement. Similarly, using different sizes or types of particles would also contribute to answering this question. Different size particles would shift the $q R$ range accessed by SAXS and hence the scale of the structures investigated with respect to the particle size. Different types of particles with a more monodisperse size distribution would interact differently with each other, possibly conforming more closely to one of the standard inter-particle potentials, which would either allow more accurate modelling of the frozen structure or confirm that the freezing process imposes a unique structure among the particles. Alternately, to focus on the structural evolution over time, experiments could be performed in which the temperature was held constant.

Finally, three other x-ray scattering techniques can provide complementary information about the samples and should be utilized for studying solidifying colloidal suspensions. First, dynamic x-ray scattering, or x-ray photon correlation spectroscopy (XPCS), allows for determination of how the particles are moving, such as distinguishing between diffusive and ballistic motion and measuring the rate of this motion. We have applied XPCS to examine particles in frozen samples and present those results elsewhere \cite{spannuthdspreprint}. A second technique is x-ray near field scattering (XNFS), which combines aspects of x-ray scattering and radiography, and also provides structural and dynamic information, though the analysis of the data is more complicated than in SAXS or XPCS \cite{cerbino2008}. However, XNFS has the benefits of accessing smaller wavevectors than SAXS and permitting observation during freezing. Third, ultra-small angle x-ray scattering (USAXS) can also access smaller wavevectors, but with a data analysis procedure similar to that for standard SAXS \cite{morvan1994}. USAXS could clarify the structure at intermediate length scales and possibly identify the length scale associated with the low-$q$ intensity upturn seen in our experiments. By combining SAXS and other x-ray techniques future work will greatly increase our knowledge of the small scale structure resulting from solidification of colloidal suspensions, which in turn will help enhance understanding of the processes occurring during solidification and allow for better control of the final solidified product.

\begin{acknowledgments}
We thank S. Narayanan, A. Sandy and M. Sprung for assistance with the SAXS experiments, and X. Lu, J. Neufeld, E. Thomson and L. Wilen for useful discussions. MS acknowledges the National Science Foundation (NSF) Graduate Research Fellowship Program for support. SGJM thanks the NSF for support via DMR-0906697. SSLP acknowledges support from the King Abdullah University of Science and Technology (KAUST), Award No. KUK-C1-013-04. JSW acknowledges support from the NSF Grant No. OPP0440841 and the US Department of Energy Grant No. DE-FG02-05ER15741. Use of the Advanced Photon Source at Argonne National Laboratory was supported by the US Department of Energy, Office of Science, Office of Basic Energy Sciences, under Contract No. DE-AC02-06CH11357.
\end{acknowledgments}

% **** X-RAY BACKGROUND **** %

\appendix*

\section{X-ray Scattering Background}
\label{xraybg}

For optically opaque materials, x-ray scattering can provide information about the structure at length scales on the order of several to around $1000$ nm. Small Angle x-ray Scattering (SAXS) probes variations in the density of electrons in a material (usually analogous to the mass density), so in colloidal suspensions SAXS data reflect the density variations associated with the size of the colloidal particles and the predominant inter-particle spacings \cite{pusey2002}. However, these density variations are disclosed in reciprocal, or Fourier, space and a model is needed to interpret the experimental results in terms of actual structure. At the most basic level though, scattering vectors (or wavevectors) with higher scattered intensity indicate structure existing within the sample on length scales proportional to the inverse of those scattering vectors. In this way, SAXS provides structural information about complex materials.

The primary result of SAXS is the scattered intensity $I \left(\mathbf{q}\right)$, where the scattering vector $\mathbf{q}$ is the vector difference between the wavevectors of the incident and the scattered x-rays. It has magnitude $q$ given by $4 \pi / \lambda \sin \left( \Theta / 2 \right)$ ($\Theta$ is the angle between the incident and scattered radiation) \cite{pusey2002}. Frequently in experiments on colloidal suspensions the scattering is expected to be isotropic, so analysis solely in terms of the magnitude of the scattering vector is acceptable. The intensity as a function of $q$ for identical particles can be expressed as
\begin{equation}
I\left(q\right) = \Phi_i \, E_{det} \, \Delta \Omega \, A_{det} \, d \, \Tr \, \phi \, V_{part} \, \left( \Delta \rho \right)^2 \, P\left(q\right) \, S\left(q\right)
\label{fullinteqn}
\end{equation}
where $\Phi_i$ is the incident x-ray flux, $E_{det}$ is the detector efficiency, $\Delta \Omega$ is the solid angle subtended by the detector, $A_{det}$ is the area of the detector, $d$ is the sample thickness, $\Tr$ is the transmission coefficient, $\phi$ is the particle volume fraction, $V_{part}$ is the volume of a single particle, $\Delta \rho$ is the electron density contrast between the particles and the solvent, $P\left(q\right)$ is the form factor, and $S\left(q\right)$ is the structure factor \cite{berneandpacora,pusey2002}. All information about the sample structure is contained within $P\left(q\right)$ and $S\left(q\right)$.

The form factor $P\left(q\right)$ describes the scattering from particles of a given size and shape. It can be calculated based on the distribution of mass within the particles for a variety of shapes (c.f. reference \cite{pedersen2002}). The structure factor $S\left(q\right)$ describes the scattering from spatial correlations among the particle positions \cite{klein2002}. It is the Fourier transform of the radial distribution function, which describes the probability of finding two particles separated by a given distance. Theoretical estimates of structure factors typically rely on radial distribution functions derived for a specified inter-particle interaction, usually pair potentials such as hard spheres or a square well (c.f. reference \cite{pedersen2002}).

For systems without an a priori model, SAXS data still provide useful information. As the Fourier transform of the mass distribution within the sample, peaks in SAXS data correspond to structures on length scales of roughly $2 \pi / q_{peak}$ \cite{pusey2002,wiese1991}. The peak width is related to the variation of this length scale around the primary one with wider peaks corresponding to greater variation. In terms of the particles' radial distribution function, the position of the primary peak in $S\left(q\right)$ represents the average distance to a particle's nearest neighbors and the height represents the average number of neighbors. Thus, SAXS provides a way to probe the structure of complex, optically opaque materials.

%\bibliography{biblio}
%

\end{document}